\documentclass{ifacconf}

\usepackage{graphicx}      
\usepackage{natbib}        

 \usepackage{amsmath} 
 \usepackage{amssymb} 
\newtheorem{Defn}{Definition}
\newtheorem{remark}{\textbf{Remark}}

\begin{document}
\begin{frontmatter}

\title{Stationary Cost Nodes in Infinite Horizon LQG-GMFGs \thanksref{footnoteinfo}} 

\thanks[footnoteinfo]{This work is supported in part by NSERC (Canada) grant RGPIN-2019-05336, the U.S. ARL and ARO grant W911NF1910110, and the U.S. AFOSR grant FA9550-19-1-0138.}

 \author[First]{Rinel Foguen Tchuendom} 
 \author[First]{Shuang Gao} 
 \author[First]{Peter E. Caines}
\address[First]{McGill University, Montreal, QC, Canada \\ (e-mail: \{fogurine,~sgao,~peterc\}@cim.mcgill.ca)}

   
   
\begin{abstract}

An analysis of infinite
 horizon linear quadratic 
Gaussian (LQG) Mean Field Games is given within the general framework of Graphon Mean Field Games (GMFG) on dense infinite graphs (or networks) introduced in  \cite{CainesHuangCDC18}. For a class of LQG-GMFGs, analytical expressions are derived for  the infinite horizon Nash 
values at the nodes of the infinite graph. Furthermore, under specific conditions on the network and the initial population means, it is shown 
that the nodes with strict local maximal infinite network degree are also nodes with strict local minimal cost at equilibrium.
\end{abstract}


\end{frontmatter}

\section{INTRODUCTION}


 This paper adds to the literature on Graphon Mean Field Games which read as Mean Field Games with networked agents, see for example  \cite{CainesHuangCDC18,H-C,LackerSoret,DelarueER,PariseAsu,CarmonaLauriere,aurell2022stochastic}. 
 These Graphon Mean Field Games are generalizations of Mean Field Games (see for example \cite{PeterEncycloMFGs,PeterSiamNews}), and can be seen as Graphon Mean Field Games with agents on large undirected graphs. To the best knowledge of the authors, this work is the first to study infinite horizon GMFGs. 
Mean field games with cost localities similar to the present paper appeared in \cite{huang2010nce}. The differences are that  in \cite{huang2010nce} each node is not assumed to be associated with an infinite mass of agents and graphons are not employed. The current paper focuses on establishing explicit analytical results on the cost at equilibrium and characterizing nodes with strict minimal (or strict maximal)  cost at equilibrium via graphon properties (like the degree of the nodes) and appropriate choice of initial conditions. To characterize nodes with strict minimal (or strict maximal) equilibrium cost, we use first and second order differential condition. 

Graphon Mean Field Games are asymptotic versions of finite large population games, with $N$ agents  ${\cal{A}}_{i}, 1 \le i \le N <  \infty$, which are distributed over the finite network, represented by its adjacency matrix $ ( g^n_{i,j} )_{i,j=1:n}$, with $n$ nodes.  We assume that, at each node $l \in \{ 1,...,n \}$ of this network, there is a cluster of agents denoted $C_l$, and let ${\bf x}_{ n}=\bigoplus_{l=1}^{n} \{x^i|i\in C_l\}$ denote the states of all agents in the population game.
Hence, the total number of agents is
$  N=\sum_{l=1}^{n} |C_l| . $  

For each agent ${\cal A}_i$ in cluster $C_i$, the coupling term (also called global mean field) governing its interaction via the network with other players, is given by: 
$$
 z^{i,n}_t=\frac{1}{n}
\sum_{l=1}^{n} g^n_{i,l} \frac{1}{|C_l|}
\sum_{j\in C_l} x^j_t,  \ \ t \geq 0, \ \forall i =1:N  .
$$
The flow of global mean fields $\{  z^{i,n}_t \  t\in [0,\infty), \ i =1:n \}$ relies on the sectional information $g^n_{i,\bullet}$ 
which represents the view of the network interactions from agents in cluster $C_l, \ l \in \{1,...,n \}$. From the point of view of an agent ${\cal A}_i$, all individuals residing in cluster $C_l$ are symmetric and their average generates an overall impact of that cluster. 

Consider the state evolution of the collection of $N$ agents  ${\cal{A}}_{i}, 1 \le i \le N <  \infty$, specified by the set of $N$ controlled linear stochastic differential equations (SDEs) over an  infinite horizon below. For each agent ${\cal{A}}_{i}$, its state denoted $x^i( \cdot ) \in \mathbb{R}$ evolves according to the SDE : 
\begin{equation}\label{NLMinorDynamics}
dx^i_t  =  b u^i_t  dt + \sigma dw^i_t , \ \forall t \geq 0, 
 \end{equation}
where $u^i( \cdot ) \in \mathbb{R}$  denotes the agent's ${\cal{A}}_{i}$ control input. For simplicity, we assume that the initial state of agent ${\cal{A}}_{i}$ is $ \ x^i_0 \sim \mathcal{N}(m^l, \nu^2),$ whenever ${\cal{A}}_{i}$ lies in cluster $C_l$, $l \in \{ 1,..,n\}$.   Let the coefficients $ a , b, m^l \ l=1:n, \in \mathbb{R}, \nu >0 $, $r > 0$,  $\sigma \geq 0$, and $\{ w^i, i=1,...,N \}$ be a collection of independent Brownian motions defined on a probability space $(\Omega, \mathbb{F}, \mathbb{P})$ satisfying the usual conditions.

We consider a scenario where each agent ${\cal{A}}_{i}$ aims to minimize infinite horizon quadratic costs
%
\begin{equation}
\label{CNP:Minor:GenCost}
J^{N} (u^i ,u^{-i}) := \mathbb{E} \int_0^\infty e^{- \rho t} \big[ r ( u^i_t )^2  +  \big( x^i_t - z^{i,n}_t \big)^2  \big] dt ,
\end{equation}
 where  $1\le i\le N$, $\rho > 0$ and $u^{-i}$ denotes the controls of all agents other than ${\cal A}_i$.
 

 
 \begin{Defn}[Nash Equilibrium]
A collection of controls, denoted $(u^{i*}, i=1,...,N)$, is a Nash equilibrium if and only if  any unilateral deviation from $u^{i*}$ to any other control $u^i$ yields a higher cost, that is,
\begin{equation}
   J_{i}^{N} (u^{i*} ,u^{-i*}) \leq  J_{i}^{N} (u^i ,u^{-i*}), \ \forall i =1,...,N.
\end{equation}
 \end{Defn}

Finding a Nash equilibrium in networked population games gets increasingly complex as both the cluster size and the network size grow. In the situation where the network describing the interaction between the agents is uniform  (i.e. fully symmetric), the theory of Mean Field Games provides satisfactory answers to this problem (see \cite{CainesHuangMalhame} \cite{LarsyLions2}). 

For non-uniform networks, Graphon Mean Field Games model asymptotic limits of population games in the double limit, $n \to \infty$ and  $\min_{l=1:n} \vert C_l \vert \to \infty$ (observe this implies that the number of agents, denoted by  $N= \sum_{l=1}^{n} |C_l|$, goes to infinity). 

We assume that the sequence of dense graphs (characterizing the networks) denoted by $\{ ( g^n_{i,j} )_{i,j=1:n} \}_{n=1}^\infty$ converges, in the cut-metric (see \cite{lovasz2012large}), to a unique limit graphon denoted
\begin{align*}
    g : [0,1] \times [0,1] & \to [0,1] \\
    (\alpha, \beta) & \mapsto g(\alpha , \beta).
\end{align*}
Graphons are bounded symmetric Lebesgue measurable functions $g: [0,1] \times [0,1] \rightarrow [0,1]$ which can be interpreted as specifying weighted links on the set of nodes $[0,1]$ (see \cite{lovasz2012large}).

With the network interaction within a cluster being uniform, we deduce that in the infinite cluster size case, at any graphon node $\alpha \in [0,1]$, there exists a representative (or typical) agent, denoted ${\cal A}_\alpha$, whose state's evolution is given by the SDE, $\forall t \in [0,\infty),$
\begin{equation}
\label{MVNLMinorDyn2}
\begin{aligned}
 dx^\alpha_t &=  b u^\alpha_t  dt + \sigma dw^\alpha_t, \quad 
 x^\alpha_0   \sim \mathcal{N}(m^\alpha , \nu^2).  
 \end{aligned}
\end{equation}


Each representative agent ${\cal A}_\alpha$ aims at minimizing an infinite horizon quadratic cost given by 
\begin{align}
\label{Jalpha}
J(u^\alpha, z^\alpha) := \mathbb{E} \int_0^\infty e^{- \rho t} \big[ r ( u^\alpha_t )^2  +  \big( x^\alpha_t - z^{\alpha}_t \big)^2 \big] dt,
\end{align}
where $r, \rho > 0$ and, the global mean field denoted by $z^{\alpha}_t \  $, are given by,  
\begin{equation}
    z^{\alpha}_t := \int_0^1 g(\alpha, \beta) \mathbb{E}[x^\beta_t] d\beta, \ \forall t \in [0,\infty), \ \forall \alpha \in [0,1] .
\end{equation}



\section{The LQG-GMFG Problem (driftless case)}


\subsection{Infinite Horizon LQG-GMFGs }

Define the following admissible control space,
\begin{align*}
    \mathbb{A} & := \{ u : \Omega \times [0,T] \mapsto \mathbb{R}  \ \vert \  u \ \textrm{is} \   \mathbb{F}-\textrm{progressively} \\
     & \quad \quad \quad \textrm{measurable and} \  \mathbb{E}  \int_0^\infty e^{ - \rho t} \vert u(t) \vert^2 dt  < \infty \}, 
\end{align*}
and the Linear Quadratic Gaussian Graphon Mean Field Games (LQG-GMFGs) problem: 
\begin{enumerate}
    \item (Mean Field Inputs) Fix a two-parameter deterministic global flow of mean fields $ \{ z^\alpha_t, \ t \in [0, +\infty), \ \alpha \in [0,1] \} $. 
 
    \item (Control Problems) Find optimal controls, denoted by $u^{\alpha, o} := (u^{\alpha, o}_t)_{t\in[0,T]} \in \mathbb{A}$, such that 
    \begin{align}
    \label{AlphaCosts}
       & J(  u^{\alpha, o}, z^\alpha  )  = \min_{ u^\alpha \in \mathbb{A}} J( u^{\alpha}, z^\alpha  )  \\ 
        & = \min_{ u^\alpha \in \mathbb{A}} \mathbb{E}  \int_0^\infty e^{- \rho t} \big[  r \big( u^\alpha_t \big)^2 + \big(x^\alpha_t - z^{\alpha}_t \big)^2 \big] dt \notag
    \end{align}
  where $\forall t \in [0,+\infty), \forall \alpha \in [0,1]$
  \begin{align}
  \label{AlphaStates}
      dx^\alpha_t & =  b u^\alpha_t dt + \sigma d w^\alpha_t ,  ~~ x^\alpha_0 \sim \mathcal{N}(m^\alpha, \nu^2). 
  \end{align}
 \item (Consistency Conditions) Show that the optimal state trajectories $\{  x^{\alpha, o}_t, \ t \in [0,\infty) , \  \forall \alpha \in [0,1] \}$, satisfy the consistency conditions, for all $ (\alpha, t) \in [0,1]\times[0, \infty)$;
 \begin{equation}
 \label{AlphaMKV_States}
     z^\alpha_t = \int_0^1 g(\alpha, \beta) \mathbb{E}[x^{\beta, o}_t] d\beta.
 \end{equation}
\end{enumerate}

The control problems can be solved via the following approach described in  \cite{HCM_2007}. Consider the following algebraic Riccati equation: 
\begin{equation}\label{eq:AREqu}
     \frac{b^2}{r} \pi^2   +  \rho \pi =  1 .
\end{equation}
The Riccati equation has a unique positive solution
\begin{equation}\label{eq:riccati}
     \pi = \sqrt{ \frac{r^2 \rho^2}{4 b^4} +  \frac{r}{b^2} } -  \frac{  \rho  r}{2 b^2} > 0.
\end{equation}

Consider $C_b \left( [0, \infty) \right)$ the set of bounded continuous functions over the domain $[0,\infty)$.  This space endowed with the supremum norm, $ \vert x \vert_{\infty} := \sup_{t \in [0,\infty)} \vert x(t) \vert $ is a Banach space. Consider $L^2 \left( [0,1] \right)$ the set of square integrable functions on the domain $[0,1]$ with inner product $ \langle x, y \rangle = \int_0^1 x(\beta) y(\beta) d \beta . $

\begin{prop}
Assume that there exists a process $ \{ s^\alpha_t, \ \alpha \in [0,1], \ t \in [0,\infty) \} \in C_b \left( [0, \infty) \right) \times L^2 \left( [0,1] \right) $ satisfying the offset ODE
\begin{equation}
    \frac{d s^\alpha_t}{dt} = \bigg( \frac{b^2}{r} \pi + \rho \bigg) s^\alpha_t + z^\alpha_t, \ \ t \in [0,\infty).
 \end{equation}
Then, there exists an optimal control process for the infinite horizon optimal control problem above, namely, for all $\alpha \in [0,1]$,
\begin{equation}
    u^{\alpha,o}_t = - \frac{b}{r} \big( \pi x^{\alpha,o}_t + s^\alpha_t \big), \ \ t \in [0,\infty),
 \end{equation}
where the optimal state process $(x^{\alpha,o}_t)_{t \in [0,T]}$ is given by the SDE, $x^{\alpha,o}_0 \sim \mathcal{N}(m^\alpha, \nu^2)$,
\begin{align*}
\notag 
   d x^{\alpha, o}_t & = \bigg[ -  \frac{b^2}{r} \pi   x^{\alpha, o}_t - \frac{b^2}{r} s^\alpha_t \bigg] dt + \sigma dw^\alpha_t . \notag 
\end{align*}
\end{prop}

\begin{pf}
The proof is a standard application of LQG tracking control theory. See for example \cite{HCM_2007}. \qed
\end{pf}

\begin{prop}
Assume that there exists a process $ \{ q^\alpha_t, \ \alpha \in [0,1], \ t \in [0,\infty) \} \in C_b \left( [0, \infty) \right) \times L^2 \left( [0,1] \right) $ satisfying the ODE
\begin{equation}
    \frac{d q^\alpha_t}{dt} = -  \sigma^2  \pi + \frac{b^2}{r} (s^\alpha_t)^2  + \rho q^\alpha_t - (z^\alpha_t)^2 . 
\end{equation}
Then, the optimal costs are given, for all $\alpha \in [0,1]$, by
\begin{align}
    J (u^\alpha, z) &= \pi \mathbb{E}[( x^{\alpha, o}_0)^2]+ 2 s^\alpha_0 \mathbb{E}[ x^{\alpha, o}_0] + q^\alpha_0  \notag  \\
    &= \pi ( \nu^2 + (m^\alpha)^2 ) + 2 s^\alpha_0 m^\alpha + q^\alpha_0.
\end{align}
\end{prop}

\begin{pf}
The proof is also standard for LQG tracking problems. See for example \cite{HCM_2007}. \qed
\end{pf}

Once the control problems have been solved and their solutions characterized by the two propositions above, we proceed to verify the consistency condition. 

\begin{prop}
Let the assumptions of Proposition 1 be in force. The consistency conditions  (\ref{AlphaMKV_States}) are satisfied  if and only if there exists a process $\{ z^\alpha_t,\ \alpha \in [0,1], \ t \in [0,\infty) \} \in C_b \left( [0, \infty) \right) \times L^2 \left( [0,1] \right)$ determined by the ODE: 
\begin{align}
\label{const_ODEs}
    d z^\alpha_t &= \bigg[ - \frac{b^2}{r} \pi  z^\alpha_t - 
    \frac{b^2}{r} \int_0^1 g(\alpha, \beta) s^\beta_t d\beta \bigg] dt,  \\
    z^\alpha_0 &= \int_0^1 g(\alpha, \beta) m^\beta d\beta . \notag 
\end{align}
\end{prop}

\begin{pf}
The consistency conditions (\ref{AlphaMKV_States}) are in fact fixed point conditions on the optimal state processes. From Proposition 1 we have an SDE representation for these optimal states. But due  to the linearity of the problem, the existence of the fixed points is characterized in terms of the existence of solutions to ODEs \eqref{const_ODEs}.  \qed
\end{pf}



Compiling the three previous propositions, we obtain that the infinite horizon LQG-GMFGs under study is solvable with explicit costs at equilibrium, whenever there exists processes $\{ z^\alpha_t, s^\alpha_t, q^\alpha_t, \alpha \in [0,1], \ t \in [0,\infty) \} \subset C_b \left( [0, \infty) \right) \times L^2 \left( [0,1] \right)$ that are solutions to the following ODEs:
\begin{align}
\label{eq_z}
&\frac{ d z^\alpha_t }{dt} =  - \frac{b^2}{r} \pi  z^\alpha_t - 
    \frac{b^2}{r} \int_0^1 g(\alpha, \beta) s^\beta_t d\beta ,  \\ 
    \label{eq_s}
       & \frac{d s^\alpha_t}{dt}  = \Big( \frac{b^2}{r} \pi + \rho \Big) s^\alpha_t + z^\alpha_t, \\
        \label{eq_q} 
  & \frac{d q^\alpha_t}{dt} = -  \sigma^2  \pi + \frac{b^2}{r} (s^\alpha_t)^2  + \rho q^\alpha_t - (z^\alpha_t)^2 ,\\
  &  z^\alpha_0 =  \int_0^1 g(\alpha, \beta) m^\beta d\beta . \nonumber
\end{align}
The main difficulty with this result is that we don't know the steady-state information $(z^\alpha_\infty, s^\alpha_\infty, q^\alpha_\infty)$ required to solve the ODEs above. To circumvent this obstacle we apply a technique from \cite{HCM_2007} which consists in solving for $(z^\alpha_\infty, s^\alpha_\infty, q^\alpha_\infty)$ from a steady state condition in the infinite horizon, 
\begin{equation}
 0=   \frac{d z^\alpha_\infty}{dt} =   \frac{d s^\alpha_\infty}{dt} =   \frac{d q^\alpha_\infty}{dt} , \quad \forall \alpha \in [0,1].
\end{equation}

This yields the family of algebraic equations indexed by $ \alpha \in [0,1]$,   
\begin{align}
0 &=  - \frac{b^2}{r} \pi   z^\alpha_\infty - 
    \frac{b^2}{r} \int_0^1 g(\alpha, \beta) s^\beta_\infty  d\beta ,  \\
    0 &=  \Big( \frac{b^2}{r} \pi + \rho \Big) s^\alpha_\infty + z^\alpha_\infty, \\
 0 &= -  \sigma^2  \pi + \frac{b^2}{r} (s^\alpha_\infty)^2  + \rho q^\alpha_\infty - (z^\alpha_\infty)^2 . \label{eq:q-infty-s-z}
\end{align}

From the first two equations, we have
\begin{align*}
    0=& \Big( - \frac{b^2}{r} \pi  \Big)  \left[ \Big(  - \frac{b^2}{r} \pi  \Big) - \rho \right] s_\infty^\alpha   - \frac{b^2}{r} \int_0^1 g(\alpha, \beta) s^\beta_\infty  d\beta,
\end{align*}
with $ \alpha \in [0,1]$,
which is equivalent (with discrepancies on at most a set of measure zero) to
\begin{align} \label{eq:sinfty-operator}
\bigg[ &  \pi     \Big(  \frac{b^2}{r} \pi  \ + \rho \Big) I -   g \bigg] \circ s_\infty =0
\end{align}
where $(g\circ s_\infty) (\cdot) := \int_0^1 g(\cdot, \beta) s_\infty(\beta) d\beta,$ and $I$ denotes the identity operator from $L^2 \left([0,1]\right)$ to $L^2\left([0,1]\right)$.

The operator 
$\bigg[  \pi     \Big(  \frac{b^2}{r} \pi  \ + \rho \Big) I -   g \bigg] = \left[ I -   g \right]  $ is invertible if
$ 1 \in \mathbb{R} $ is not an eigenvalue of the operator $g$.

\begin{remark}
Since it is assumed that $|g(x,y)|\leq 1$, for all $x,y \in  [0,1]$,
the operator norm of $g$ satisfies $$\|g\|_\textup{op} : =\sup_{v \in L^2[0,1]}\frac{\|g \circ v\|}{\|v\|} \leq \sqrt{\int_{[0,1]^2} g(x,y)^2 dxdy} \leq 1, $$
following  $\textup{\cite[Lemma 7]{gao2020graphon}}$. 
This implies that the absolute values of all the eigenvalues of $g$ are less than or equal to $1$.
\end{remark}

\textbf{Assumption (A1)}: 
All eigenvalues of the graphon operator $g$ are strictly less than $1$.

Under {Assumption (A1)}, the functional equation \eqref{eq:sinfty-operator} admits the (unique) solution in $L^2([0,1])$ with
\begin{equation}\label{eq:terminalsz}
    z^\alpha_\infty = 0 = s^\alpha_\infty,   ~ \alpha \in [0,1],
\end{equation}
and an application of \eqref{eq:q-infty-s-z} yields
\begin{equation}\label{eq:terminalq}
    q^\alpha_\infty = \frac{ \sigma^2  \pi}{\rho},  ~ \alpha \in [0,1]. 
\end{equation}

We are interested in calculating an explicit  solution, $\{ z^\alpha_t, s^\alpha_t, q^\alpha_t, \alpha \in [0,1], \ t \in [0,\infty) \} \subset C_b \left( [0, \infty) \right) \times L^2 \left( [0,1] \right)$, to the ODEs (\ref{eq_z}-\ref{eq_s}-\ref{eq_q}) 
with the infinite horizon conditions
\begin{align}
  z^\alpha_\infty = 0 = s^\alpha_\infty, \quad q^\alpha_\infty = \frac{ \sigma^2 \pi}{\rho},  ~ \alpha \in [0,1]. 
\end{align}

\textbf{Assumption (A2)} The graphon $g$ is of finite rank, that is, there exists $L <\infty$ such that 
\[
g (\alpha,\beta) = \sum_{\ell =1}^L \lambda_\ell f_\ell(\alpha) f_\ell(\beta) ,
\]
where $f_\ell$ is the orthonormal eigenfunction associated with the non-zero eigenvalue $\lambda_\ell$ of $g$.


\begin{prop}
Let assumption (A2) be in force. Then, the process $ \{ z^\alpha_t, s^\alpha_t \ \alpha \in [0,1], \ t \in [0,\infty) \}$
is explicitly given as below $\ \forall t \geq 0, \  ~\alpha \in [0,1]$,
\begin{align}
    z^\alpha_t &= \sum_{l =1}^L  f_\ell (\alpha) z^\ell_t,   \\
    s^\alpha_t & = - \sum_{l =1}^L   f_\ell (\alpha) \bigg( \frac{z^\ell_t}{\theta(\lambda_\ell) + \theta(0)} \bigg) , \notag
\end{align}
where  for $ \ell \in \{1, \ldots, L \}$ 
\begin{align}
    z^\ell_t = \lambda_\ell \langle m, f_\ell \rangle \exp \left[  \left( \frac{\rho}{2} -\theta(\lambda_\ell) \right) t \right] 
\end{align}
and $\theta(\cdot)$ is a function defined by
\begin{align}
    \theta (\tau)  
& := \sqrt{\frac{ \rho^2}{4} +  (1-\tau) \frac{b^2}{r} },  \quad \tau \in R.
\end{align}
\end{prop}

\begin{pf}
Consider the graphon spectral decomposition under the finite rank assumption (A2),
\begin{equation}\label{eq:g-alpha-beta}
    g(\alpha, \beta) = \sum_{\ell =1}^L \lambda_\ell f_\ell(\alpha) f_\ell(\beta), \quad \forall \alpha, \beta \in [0,1] ,
\end{equation}
or equivalently written as 
\[
g = \sum_{\ell =1}^L \lambda_\ell f_\ell f_\ell^T, \quad f_\ell \in L^2\left([0,1]\right) ,
\]
where $f_\ell$ is the orthonormal eigenfucntion of $g$, and $\lambda_\ell$ is the eigenvalue associated with $f_\ell$.  By the definition of eigenvalues and eigenfuntions, 
\[
g \circ f_\ell  = \lambda_\ell f_\ell.
\]
Following the spectral reformulation of two point boundary value problems developed in \cite{ShuangRinelPeterCIS,gao2021lqg}, we define the eigen processes
\[
z_t^\ell = \langle z_t, f_\ell \rangle, \quad s_t^\ell = \langle s_t, f_\ell \rangle, \quad t\in [0,\infty], ~\ell \in \{1,2,...\}.
\]
These processes are solutions to the following equations:
\begin{align*}
    \frac{d {z}^{\ell}_t }{ dt} & =  - \frac{b^2 \pi}{r}  {z}^{\ell}_t - \lambda_\ell \frac{b^2}{r} {s}^{\ell}_t , \quad  {z}^{\ell}_0 = \lambda_\ell \langle m, f_\ell \rangle ,  \\
 \frac{d {s}^{\ell}_t }{ dt} & = {z}^{\ell}_t  + \bigg(    \frac{b^2 \pi}{r} +\rho \bigg) {s}^{\ell}_t , \quad {s}^{\ell}_\infty = 0,
\end{align*}
from which we seek an explicit solution that is compatible with the infinite horizon condition ${z}^{\ell}_\infty = 0$, for all $\ell \in \{ 1, \ldots, L \}$.
From the ODE for $s^\ell$, it admits the representation below:
\begin{align}
  s^\ell_t &= - \int_t^\infty \exp \left( \left( \frac{b^2 \pi}{r} + \rho \right) (t-s) \right) z^\ell_s ds, \\
  &= - \int_t^\infty \exp \left( \frac{1}{\pi} (t-s) \right) z^\ell_s ds, 
\end{align}

which is derived invoking the Riccati equation (\ref{eq:AREqu}),
\begin{align}
\left( \frac{b^2 \pi}{r} + \rho \right) =  \frac{1}{\pi} \ ,
\end{align}
and which satisfies the condition 
\begin{align}
     s^\ell_\infty = 0 .
\end{align}
By substituting this expression for $s^\ell$ back into the ODE for $z^\ell$, we obtain the representation below
\begin{align}
   &  \frac{d  z^\ell_t}{dt} = - \frac{b^2 \pi }{r}  z^\ell_t   + \lambda_\ell \frac{b^2}{r}\int_t^\infty \exp \left(  \frac{1}{\pi} (t-s) \right) z^\ell_s ds  \notag  .
\end{align}
By differentiating the above ODE and making appropriate substitutions, we obtain the second order ODE for $z^\ell$,
\begin{align}
   &  \frac{d^2 z^\ell_t}{dt} - \rho \frac{d z^\ell_t}{dt}  + \frac{b^2}{r} \big( \lambda_\ell  - 1 \big) z^\ell_t = 0. \notag
\end{align}
This can be solved via a characteristic equation 
\begin{align}
 &  \xi_\ell^2   - \rho  \xi_\ell    + \frac{b^2}{r} \big( \lambda_\ell  - 1 \big) = 0, \notag 
\end{align}
which admits as a negative solution
\begin{align}
    \xi_\ell & =  \left( \frac{\rho}{2}  - \sqrt{ \frac{\rho^2}{4}  +  \frac{b^2}{r} \left( 1  - \lambda_\ell \right)  } \  \right)  \notag  =   \frac{\rho}{2} -\theta(\lambda_\ell) < 0,
\end{align}
under Assumption (A1).
%
%
%
We thus obtain 
\begin{align}
    z^\ell_t = \lambda_\ell \langle m, f_\ell \rangle \exp \left( \xi_\ell t \right) ,  \ \forall t \geq 0, 
\end{align}

where, because $\xi_\ell < 0$ for all $l \in \{1, \ldots, L \}$, the infinite horizon condition $z^\ell_\infty = 0$ is satisfied. 

We now proceed to calculate  $s^\ell$ as below, $\forall t \in [0,\infty)$
\begin{align}
     s^\ell_t &= - \lambda_\ell \langle m, f_\ell \rangle \exp \left( \frac{t}{\pi} \right)  \int_t^\infty \exp \left( \left( \xi_\ell - \frac{1}{\pi} \right)s \right) ds, \notag 
\end{align}
then
   $\left( \xi_\ell - \frac{1}{\pi} \right) < 0 $
gives
\begin{align}
   s^\ell_t  & = \lambda_\ell \langle m, f_\ell \rangle   \left( \xi_\ell - \frac{1}{\pi} \right)^{-1} \exp \left( \xi_\ell t \right)  .
\end{align}
Also, for all $\ell \in \{ 1, \ldots, L \}$, we have that 
\begin{align*}
    \xi_\ell & - \frac{1}{\pi}  = \frac{\rho}{2} - \theta(\lambda_\ell) - \frac{1}{\pi}  \\
    & = - \theta(\lambda_\ell) - \left( \frac{\rho^2}{4} + \frac{b^2}{r} \right)^\frac{1}{2}  = - \theta(\lambda_\ell) - \theta(0).
\end{align*}
Therefore, it holds that
\begin{align}
   s^\ell_t  & = - \frac{z^\ell_t}{  \theta(\lambda_\ell) + \theta(0)  }, \quad \forall \ell \in \{ 1, \ldots, L \}.
\end{align}

Based on \eqref{eq:g-alpha-beta} and the definition of the eigen processes, we can now reconstruct the solution $ \{ z^\alpha_t, s^\alpha_t\ \alpha \in [0,1], \ t \in [0,\infty) \}$ as below $ \forall t \geq 0,   ~\alpha \in [0,1]$,
\begin{align}
    z^\alpha_t &= \sum_{l =1}^L  f_\ell (\alpha) z^\ell_t,   \quad
    s^\alpha_t = - \sum_{l =1}^L f_\ell (\alpha)  \frac{z^\ell_t}{ \theta(\lambda_\ell) + \theta(0) }  . \qed \notag
\end{align}

\end{pf}

\begin{prop} \label{prop:cost-form}
Let the assumptions (A1)-(A2) be in force. Then the cost at equilibrium is explicitly given, for every $ \alpha \in [0,1]$, below
\begin{align*}
       J (u^\alpha, z) = \pi & \nu^2 + \pi (m^\alpha)^2 + \frac{ \sigma^2 \pi}{\rho}  - 2 m^\alpha \sum_{l =1}^L   f_\ell (\alpha)  \bar{\lambda}_\ell \langle m, f_\ell \rangle \\
      +  \frac{1}{\rho} \bigg(  \sum_{\ell = 1}^L &  f_\ell  (\alpha) \lambda_\ell \langle m, f_\ell \rangle \bigg)^2 
    - \frac{b^2}{r \rho }  \left( \sum_{\ell = 1}^L  f_\ell (\alpha) \bar{\lambda}_\ell \langle m, f_\ell \rangle \right)^2 \\
     - \sum_{k=1}^L \sum_{\ell = 1}^L & f_k (\alpha) f_\ell (\alpha)  \langle m, f_k \rangle \langle m, f_\ell \rangle \left( \frac{\rho}{2} - \theta(\lambda_k) \right)  \\
    & \quad \quad    \left( \frac{1}{\theta(\lambda_\ell) + \theta(\lambda_k) }  \right)  \left[  \frac{2}{\rho} \lambda_k \lambda_\ell - \frac{2 b^2}{\rho r} \bar{\lambda}_k \bar{\lambda}_\ell  \right],
\end{align*}
where we define
\begin{align} \label{eq:lambda-bar2}
    \bar{\lambda}_\ell := \frac{\lambda_\ell}{\theta(\lambda_\ell) + \theta(0)}, \quad \ell \in \{1, \ldots, L \} .
\end{align}
\end{prop}

\begin{pf}
Given the process $ \{ z^\alpha_t, s^\alpha_t \ \alpha \in [0,1], \ t \in [0,\infty) \}$ explicitly calculated for every $\alpha \in [0,1]$, we proceed to calculate explicitly the process $ \{ q^\alpha_t, \ \alpha \in [0,1], \ t \in [0,\infty) \}$, for every $\alpha \in [0,1]$. 

A straightforward calculation allows one to verify that
\begin{equation}\label{eq:qt-explicit}
    q^\alpha_t = - \exp\left( \rho t \right) \int_t^\infty \Theta(\alpha, s) \exp\left( - \rho s \right) ds, 
\end{equation}
with $\Theta(\alpha, t), \ \forall \alpha \in [0,1], \ t \in [0,\infty), $  defined by:
\begin{align*}
   &  \Theta(\alpha,t)   = -  \sigma^2  \pi  - \left( z^\alpha_t \right)^2 + \frac{b^2}{r} (s^\alpha_t)^2 ,  \\
   &  \quad \quad =  - \sigma^2 \pi  -  \left( \sum_{l=1}^L  f_\ell (\alpha) \lambda_\ell \langle m, f_\ell \rangle  \exp \left( \xi_\ell t \right) \right)^2 \\
   & +\frac{b^2}{r}  \left(  \sum_{\ell =1}^L  \left( \theta(\lambda_\ell) + \theta(0) \right)^{-1}  f_\ell (\alpha) \lambda_\ell \langle m, f_\ell \rangle  \exp \left( \xi_\ell t \right)  \right)^{2} , \\
    &  \quad \quad =  - \sigma^2 \pi  -  \left( \sum_{l=1}^L  f_\ell (\alpha) \lambda_\ell \langle m, f_\ell \rangle  \exp \left( \xi_\ell t \right) \right)^2 \\
& \quad \quad  +\frac{b^2}{r}  \left(  \sum_{\ell =1}^L   f_\ell (\alpha) \bar{\lambda}_\ell \langle m, f_\ell \rangle  \exp \left( \xi_\ell t \right)  \right)^{2} ,
\end{align*}

is a solution to the offset ODE, 
\begin{equation}
     \frac{d q^\alpha_t}{dt} = -  \sigma^2  \pi + \frac{b^2}{r} (s^\alpha_t)^2  + \rho q^\alpha_t -  \left( z^\alpha_t \right)^2 .
\end{equation}

Moreover, the process $\{ q^\alpha_t , \  \alpha \in [0,1], \ t \in [0,\infty)\} $ is compatible with the infinite horizon condition
\begin{align}\label{eq:lambda_bar}
    q^\alpha_\infty = \frac{ \sigma^2  \pi}{\rho}.
\end{align}

Indeed, by applying L'Hopital's Rule, we obtain that,
\begin{align*}
    \lim_{t \to \infty} & q^\alpha_t = \lim_{t \to \infty}  - \exp\left( \rho t \right) \int_t^\infty \Theta(\alpha, s) \exp\left( - \rho s \right) ds \\
    &= \lim_{t \to \infty} \frac{\Theta(\alpha, t)}{ - \rho} = \frac{- \sigma^2  \pi}{- \rho} = q^\alpha_\infty.
\end{align*}

Recall that the cost at equilibrium is given, for every $ \alpha \in [0,1]$, by 
\begin{align}
\label{xx}
     & J (u^\alpha, z) = \pi ( \nu^2 + (m^\alpha)^2 ) + 2 s^\alpha_0 m^\alpha + q^\alpha_0.
\end{align}

Therefore, to calculate the cost explicitly, for all $\alpha \in [0,1]$, it is enough to calculate the quantities $s^\alpha_0, q^\alpha_0$. 

Recall that for every $\alpha \in [0,1],$
\begin{align*}
s^\alpha_0 & = - \sum_{l =1}^L   f_\ell (\alpha)  \bar{\lambda}_\ell \langle m, f_\ell \rangle, \  q^\alpha_0 = -  \int_0^\infty \Theta(\alpha, s) \exp\left( - \rho s \right) ds, 
\end{align*}

where $\Theta(\alpha, t), \ \forall \alpha \in [0,1], \ t \in [0,\infty), $ is defined by:
\begin{align*}
   &  \Theta(\alpha,t)  = - \sigma^2 \pi  -  \left( \sum_{l=1}^L  f_\ell (\alpha) \lambda_\ell \langle m, f_\ell \rangle  \exp \left( \xi_\ell t \right) \right)^2 \\
   & \quad \quad +\frac{b^2}{r}  \left(  \sum_{\ell =1}^L   f_\ell (\alpha) \bar{\lambda}_\ell \langle m, f_\ell \rangle  \exp \left( \xi_\ell t \right)  \right)^{2}.
\end{align*}

Integrating by parts yields
\begin{align*}
    q^\alpha_0 = - \frac{\Theta(\alpha,0)}{\rho} - \frac{1}{\rho} \int_0^\infty \exp\left( - \rho s \right) \frac{d \Theta(\alpha,s)}{ds} ds.
\end{align*}
We then calculate that, 
\begin{align*}
  - \frac{ \Theta(\alpha, 0)}{\rho} & = \frac{ \sigma^2 \pi}{\rho}  + \frac{1}{\rho} \left( \sum_{\ell = 1}^L f_\ell (\alpha) \lambda_\ell \langle m, f_\ell \rangle \right)^2  \\
    - \frac{b^2}{r \rho } & \left( \sum_{\ell = 1}^L  f_\ell (\alpha) \bar{\lambda}_\ell \langle m, f_\ell \rangle \right)^2 ,
\end{align*}
and
\begin{align*}
 - & \frac{1}{\rho}  \int_0^\infty \exp\left( - \rho s \right) \frac{d \Theta(\alpha,s)}{ds} ds \\
 & = \sum_{k=1}^L \sum_{\ell = 1}^L \xi_k \left( \int_0^\infty e^{\left( \xi_k + \xi_\ell - \rho  \right) s} ds \right) \\
  & \quad \quad \quad f_k (\alpha) f_\ell (\alpha)  \langle m, f_k \rangle \langle m, f_\ell \rangle  \left[  \frac{2}{\rho} \lambda_k \lambda_\ell - \frac{2 b^2}{\rho r} \bar{\lambda}_k \bar{\lambda}_\ell  \right]\\
  & = \sum_{k=1}^L \sum_{\ell = 1}^L \xi_k  \left( \xi_k + \xi_\ell - \rho  \right)^{-1}  f_k (\alpha) f_\ell (\alpha)  \langle m, f_k \rangle \langle m, f_\ell \rangle \\
   & \hspace{2cm} \left[  \frac{2}{\rho} \lambda_k \lambda_\ell - \frac{2 b^2}{\rho r} \bar{\lambda}_k \bar{\lambda}_\ell  \right] .
\end{align*}
Therefore, by observing the equality
\begin{align*}
     \left( \xi_k + \xi_\ell - \rho  \right) = - \left( \theta(\lambda_\ell) + \theta(\lambda_k) \right),
\end{align*}
we deduce that,
\begin{align*}
    q^\alpha_0 &=  \frac{1}{\rho} \left( \sum_{\ell = 1}^L f_\ell (\alpha) \lambda_\ell \langle m, f_\ell \rangle \right)^2 
    - \frac{b^2}{r \rho }  \left( \sum_{\ell = 1}^L  f_\ell (\alpha) \bar{\lambda}_\ell \langle m, f_\ell \rangle \right)^2 \\
    & - \sum_{k=1}^L \sum_{\ell = 1}^L \left( \frac{\rho}{2} - \theta(\lambda_k) \right)  \left( \theta(\lambda_\ell) + \theta(\lambda_k) \right)^{-1}  \\
     \quad & \quad f_k (\alpha) f_\ell (\alpha)  \langle m, f_k \rangle \langle m, f_\ell \rangle  \left[  \frac{2}{\rho} \lambda_k \lambda_\ell - \frac{2 b^2}{\rho r} \bar{\lambda}_k \bar{\lambda}_\ell  \right] + \frac{ \sigma^2 \pi}{\rho} .
\end{align*}
Finally, recalling that the cost at equilibrium is explicitly given by (\ref{xx}) and substituting the calculated terms appropriately, we get that for every $ \alpha \in [0,1]$,
\begin{align*}
     J (u^\alpha, z) = \pi & \nu^2 + \pi (m^\alpha)^2 + \frac{ \sigma^2 \pi}{\rho}  - 2 m^\alpha \sum_{l =1}^L   f_\ell (\alpha)  \bar{\lambda}_\ell \langle m, f_\ell \rangle \\
    +  \frac{1}{\rho} \bigg(  \sum_{\ell = 1}^L & f_\ell (\alpha) \lambda_\ell \langle m, f_\ell \rangle \bigg)^2 
   - \frac{b^2}{r \rho }  \left( \sum_{\ell = 1}^L  f_\ell (\alpha) \bar{\lambda}_\ell \langle m, f_\ell \rangle \right)^2 \\
     - \sum_{k=1}^L \sum_{\ell = 1}^L & f_k (\alpha) f_\ell (\alpha)  \langle m, f_k \rangle \langle m, f_\ell \rangle \left( \frac{\rho}{2} - \theta(\lambda_k) \right)  \\
    & \quad \quad    \left( \frac{1}{\theta(\lambda_\ell) + \theta(\lambda_k) }  \right)  \left[  \frac{2}{\rho} \lambda_k \lambda_\ell - \frac{2 b^2}{\rho r} \bar{\lambda}_k \bar{\lambda}_\ell  \right].
\end{align*} 
\qed
\end{pf}

\begin{prop}[Simplifications] 
\label{simplifiedcost}
Assume (A1)-(A2) hold. Then, the cost at equilibrium is explicitly given, for every $ \alpha \in [0,1]$, below
{
\begin{align*}
      J (u^\alpha, z) &= \pi \nu^2 + \pi (m^\alpha)^2 + \frac{ \sigma^2 \pi}{\rho}  - 2 m^\alpha \sum_{l =1}^L   f_\ell (\alpha)  \bar{\lambda}_\ell \langle m, f_\ell \rangle \\
   - \sum_{k=1}^L \sum_{\ell = 1}^L & f_k (\alpha) f_\ell (\alpha)  \langle m, f_k \rangle \langle m, f_\ell \rangle  \\
    & \quad \quad    \left( \frac{  {\rho} }{\theta(\lambda_\ell) + \theta(\lambda_k) } -2 \right)  \left[  \frac{1}{\rho} \lambda_k \lambda_\ell - \frac{ b^2}{\rho r} \bar{\lambda}_k \bar{\lambda}_\ell  \right] .
\end{align*}}
with $\bar{\lambda}_\ell$, $\ell \in\{1,..., L\}$, defined in \eqref{eq:lambda-bar2}.
\end{prop}
\begin{pf}
Observe that
\[
\begin{aligned}
     \sum_{k=1}^L \sum_{\ell = 1}^L & f_k (\alpha) f_\ell (\alpha)  \langle m, f_k \rangle \langle m, f_\ell \rangle  \\
    & \quad \quad    \left( \frac{  {\rho} }{\theta(\lambda_\ell) + \theta(\lambda_k) }  \right)  \left[  \frac{1}{\rho} \lambda_k \lambda_\ell - \frac{ b^2}{\rho r} \bar{\lambda}_k \bar{\lambda}_\ell  \right]\\
    -  \frac{1}{\rho}  \bigg( \sum_{\ell = 1}^L & f_\ell (\alpha) \lambda_\ell \langle m, f_\ell \rangle \bigg)^2 
    + \frac{b^2}{r \rho }  \left( \sum_{\ell = 1}^L  f_\ell (\alpha) \bar{\lambda}_\ell \langle m, f_\ell \rangle \right)^2 \\
   =  \sum_{k=1}^L \sum_{\ell = 1}^L & f_k (\alpha) f_\ell (\alpha)  \langle m, f_k \rangle \langle m, f_\ell \rangle \left( \frac{\rho}{2} - \theta(\lambda_k) \right)  \\
    & \quad \quad    \left( \frac{1}{\theta(\lambda_\ell) + \theta(\lambda_k) }  \right)  \left[  \frac{2}{\rho} \lambda_k \lambda_\ell - \frac{2 b^2}{\rho r} \bar{\lambda}_k \bar{\lambda}_\ell  \right].
\end{aligned}
\]
Taking the cost form in Prop. \ref{prop:cost-form}, then last three terms there can be further simplified, which completes the proof. \qed
\end{pf}


\section{Stationary Costs and Maximal Degree Nodes}

\textbf{Assumption (A3)}
The initial means are constant across all nodes, that is, for all $\alpha \in [0,1]$,
\begin{align}
m^\alpha = m, \quad \textrm{for some} \ \  m \neq 0,~ m \in R.
\end{align}

\begin{prop}
\label{constantmeancost}
Assume that assumptions (A1)-(A2)-(A3) hold. The equilibrium costs admit the following representation, for  every $\alpha \in [0,1]$,
\begin{align*}
      J (u^\alpha, z) &= \pi \left( \nu^2 +  m^2 + \frac{ \sigma^2 }{\rho} \right) - 2 m^2 \int_0^1 \bar{g}(\alpha, \beta) d\beta  \\
    & \hspace{2cm} - m^2 \int_0^1 \tilde{g}(\alpha, \beta \ \vert \alpha ) d\beta,
\end{align*}
where the introduced finite-rank  graphons $ \bar{g}(\cdot, \cdot)$ and  $\tilde{g}(\cdot, \cdot \ \vert \ \alpha)$ with $\alpha \in [0,1]$ are defined for all $(\epsilon, \beta) \in [0,1] \times [0,1]$  by 
\begin{align}
    \bar{g}(\epsilon, \beta ) & := \sum_{k=1}^L \bar{\lambda}_k f_k (\epsilon) f_k (\beta), \\
    \tilde{g}( \epsilon, \beta \ \vert \alpha ) & := \sum_{k=1}^L \tilde{\lambda}_k^\alpha  f_k (\epsilon) f_k (\beta),
\end{align}
and for all $k \in \{1, \ldots, L \}$, for all $\alpha \in [0,1]$, the eigenvalues are defined by
\begin{align}
    \bar{\lambda}_k = \frac{\lambda_k}{\theta(\lambda_k) + \theta(0) }, \notag 
\end{align}
\begin{align*}
 \tilde{\lambda}_k^\alpha &:=    \sum_{\ell =1}^L f_\ell (\alpha) \langle 1, f_\ell \rangle  \left( \frac{\rho}{\theta(\lambda_\ell) + \theta(\lambda_k) } - 2 \right) \\
 & \hspace{2cm} \left( \frac{1}{\rho} \lambda_k \lambda_\ell - \frac{ b^2}{\rho r} \bar{\lambda}_k \bar{\lambda}_\ell  \right) .
\end{align*}

\end{prop}

\begin{pf}
From assumptions (A1)-(A2) and Proposition \ref{simplifiedcost}, we have that the equilibrium cost is given, for every $\alpha \in [0,1]$, by
\begin{align*}
  J (u^\alpha, z) & = \pi \nu^2 + \pi (m^\alpha)^2 + \frac{ \sigma^2 \pi}{\rho}  - 2 m^\alpha \sum_{l =1}^L   f_\ell (\alpha)  \bar{\lambda}_\ell \langle m, f_\ell \rangle \\
  & \quad- \sum_{k=1}^L \sum_{\ell = 1}^L f_k (\alpha) f_\ell (\alpha)  \langle m, f_k \rangle \langle m, f_\ell \rangle  \\
    & ~ \quad    \left( \frac{  {\rho} }{\theta(\lambda_\ell) + \theta(\lambda_k) } -2 \right)  \left[  \frac{1}{\rho} \lambda_k \lambda_\ell - \frac{ b^2}{\rho r} \bar{\lambda}_k \bar{\lambda}_\ell  \right] .
\end{align*}
Assuming that (A3) hold, we get
\begin{align*}
   J (u^\alpha, z) &= \pi \left( \nu^2 +  m^2 + \frac{ \sigma^2}{\rho} \right) - 2 m^2 \sum_{l =1}^L   f_\ell (\alpha)  \bar{\lambda}_\ell \langle 1, f_\ell \rangle \\
    & \hspace{2cm} - m^2 \sum_{k=1}^L \tilde{\lambda}^\alpha_k f_k (\alpha)  \langle 1, f_k \rangle.
\end{align*}
Interpreting the quantities $\bar{\lambda}_k, \ \tilde{\lambda}^\alpha_k$, for all $\alpha \in [0,1]$  as eigenvalues, we deduce that the  cost can be written as a function of newly introduced graphons $\bar{g}(\cdot, \cdot)$ and $\tilde{g}(\cdot, \cdot)$ built from the original graphon $g(\cdot, \cdot)$. 
\hspace{2cm} \qed
\end{pf}

\textbf{Assumption (A4)} Assume that the orthonormal eigenfunctions $\{f_\ell\}_{\ell=1}^L$ of $g$ satisfy the condition
\begin{align}
    \langle 1, f_\ell \rangle =  \int_0^1 f_\ell (\beta) d\beta   < 0, \quad  \forall \ell \in \{ 1,...,L \}. 
\end{align}

Given any orthonormal eigenfunction set $\{f_\ell\}_{\ell=1}^L$ of graphon $g$, we can construct $\{\tilde{f}_\ell\}_{\ell=1}^L$ as follows:
\[
\tilde{f}_\ell = \begin{cases}
{f}_\ell, &  \text{if}~ \langle 1, f_\ell \rangle<0, \\
-{f}_\ell, &  \text{if}~ \langle 1, f_\ell \rangle\geq 0,
\end{cases} \qquad \forall \ell \in \{1,...,L\}.
\]
Then  $\{\tilde{f}_\ell\}_{\ell=1}^L$ is also an orthonormal eigenfunction set of the same graphon $g$ and it satisfies  $\langle 1, \tilde{f}_\ell \rangle\leq 0$ for any $\ell\in\{1,..., L\}$.  Thus, if all the eigenfunctions (associated with nonzero eigenvalues) of $g$ are not orthogonal to $1 \in L^2([0,1])$, there is always a choice of orthonormal eigenfunctions of $g$ that satisfies (A4); in other words, the only restriction that (A4) poses on the graphon $g$ is that all its eigenfunctions (associated with nonzero eigenvalues) should not be orthogonal to $1\in L^2([0,1])$.


\textbf{Assumption (A5)} Assume that the eigenvalues $\lambda_\ell$ of $g$, satisfy, for all $\ell, k \in \{1,...,L \}$,
\begin{equation}
\theta(\lambda_\ell) + \theta(\lambda_k)  = \frac{\rho}{2} .
\end{equation}

Given a graphon $g$ and some node $\alpha \in [0,1]$, the degree of node $\alpha \in [0,1]$, is defined by
\begin{align}
  \delta(\alpha) :=  \int_0^1 g(\alpha, \beta) d \beta = \sum_{\ell = 1}^L \lambda_\ell \langle 1, f_\ell \rangle f_\ell (\alpha) . 
\end{align}


\begin{prop}
\label{mincostmindeg}
Let the assumptions (A1) to (A5) be in force. Assume that the eigenfunctions are twice differentiable and there exists a node $\alpha^* \in (0,1)$ such that
\begin{align}
\label{prop7}
    \partial^2_{\alpha, \alpha } f_\ell ( \alpha^* ) > 0, \quad  \forall \ell \in \{ 1,...,L \}.
\end{align}

Then, the node $\alpha^* \in (0,1)$ is a node with strict local maximal degree if and only if it  is a node with strict local minimal cost. 
\end{prop}

\begin{pf}

\textit{Step 1} Assume that $\alpha^* \in [0,1]$ is  a node with strict local maximal degree. 

From assumption (A4), it follows from the first order condition 
\begin{equation}
    \partial_\alpha \int_0^1 g(\alpha^*, \beta) d\beta = \sum_{\ell = 1}^L \lambda_\ell \langle 1, f_\ell \rangle \partial_\alpha f_\ell (\alpha^*) = 0,
\end{equation}
that 
\begin{align}
\label{refpf1}
    \partial_\alpha f_\ell (\alpha^*) = 0, \quad  \forall \ell \in \{ 1,...,L \}.
\end{align}
To show that $\alpha^* \in [0,1]$ is a node with strict local minimal cost, we show that 
\begin{align}
    \partial_\alpha J (u^{\alpha^*}, z) = 0 , \quad \partial^2_{\alpha, \alpha} J (u^{\alpha^*}, z) > 0 .
\end{align}

From assumption (A5), it follows that,
\begin{align*}
    \partial_\alpha J (u^{\alpha^*}, z) & = - 2  m^2  \sum_{\ell = 1}^L \langle 1, f_\ell \rangle \bar{\lambda}_\ell \partial_\alpha f_\ell (\alpha^*) \\
    \partial^2_{\alpha, \alpha} J (u^{\alpha^*}, z) & =   - 2  m^2  \sum_{\ell = 1}^L \langle 1, f_\ell \rangle \bar{\lambda}_\ell \partial^2_{\alpha, \alpha} f_\ell (\alpha^*).
\end{align*}

So from (\ref{refpf1}), $\partial_\alpha J (u^{\alpha^*}, z) = 0$ and from (\ref{prop7}) and (A4) that $\partial^2_{\alpha, \alpha} J (u^{\alpha^*}, z) > 0 $, and hence $\alpha^*$  is  a node with  strict local minimal cost. 

\textit{Step 2} Assume that $\alpha^* \in [0,1]$ is  a node with strict local  minimal  cost. 

From assumption (A5), we compute that,
\begin{align*}
    \partial_\alpha J (u^{\alpha^*}, z) & = - 2  m^2  \sum_{\ell = 1}^L \langle 1, f_\ell \rangle \bar{\lambda}_\ell \partial_\alpha f_\ell (\alpha^*).
\end{align*}

To show that $\alpha^* \in [0,1]$ is   a node with strict local maximal degree, we show that
\begin{align}
\label{pf1}
    \partial_\alpha & \int_0^1 g(\alpha^*, \beta) d\beta = \sum_{\ell = 1}^L \lambda_\ell \langle 1, f_\ell \rangle \partial_\alpha f_\ell (\alpha^*) = 0, \\
    \label{pf2}
    \partial^2_{\alpha, \alpha} & \int_0^1 g(\alpha^*, \beta) d\beta = \sum_{\ell = 1}^L \lambda_\ell \langle 1, f_\ell \rangle \partial^2_{\alpha, \alpha} f_\ell (\alpha^*) < 0.
\end{align}

Since $\alpha^* \in [0,1]$ is  a node with strict local minimal cost, it follows that \begin{align}
    \partial_\alpha J (u^{\alpha^*}, z) & = - 2  m^2  \sum_{\ell = 1}^L \langle 1, f_\ell \rangle \bar{\lambda}_\ell \partial_\alpha f_\ell (\alpha^*) = 0,
\end{align}
and by (A4) it follows that
\begin{align}
\label{pfeq2}
    \partial_\alpha f_\ell (\alpha^*) = 0, \quad \forall \ell \{1,...,L \}. 
\end{align}

Then by (\ref{pfeq2}) and (A4) together with (\ref{prop7}), it follows that (\ref{pf1}) and (\ref{pf2}) hold.  \qed
\end{pf}

\begin{remark}
Throughout this paper, whenever  $\alpha^* \in (0,1)$ is a critical (i.e. locally maximal or minimal) interior point of   $J (u^{\alpha^*}, z)$ it is assumed that 
\begin{equation}
    \frac{\partial J (u^{\alpha^*}, z)}{\partial \alpha} =0, 
\end{equation}
which requires that differentiation with respect to  $\alpha \in (0,1)$ is meaningful within the LQG-GMFG framework introduced in \cite{CDC21CriticalNodes}.  
 To justify this, differential calculus for GMFGs with respect to node parameterization is made rigorous via the formulation of embedded vertexon graphons in  compact subsets of $R^d$, for some $d \geq 1,$ in  \cite{PeterVertexon}. The further development of this topic is a future direction of the work in the current paper.
\end{remark}

\section{Conclusion}
In this work, we explicitly solve a class of infinite horizon linear quadratic Gaussian  Graphon Mean Field Games. We show that under appropriate conditions on the graphon eigenvalues and eigenfunctions, and the initial population means, the nodes with strict local maximal degree are also nodes with strict local minimal cost at equilibrium. Although the conditions employed are quite restrictive, we make them in order to obtain a first set of results and to gain the intuition required to relax them in future work.

\bibliography{ifacconf.bib}

\end{document}